\documentclass[prb,aps,twocolumn,superscriptaddress,showpacs]{revtex4}
\usepackage[dvips]{graphicx}
\usepackage{amsmath}
\usepackage{color,bm}

\def\rr{{\bf r}}
\def\rr{{\bf r}}

\def\k{{\bf k}}

\def\rr{{\bf r}}

\def\vs{{\bf v}_s}

\begin{document}

\title{Inversion of specific heat oscillations with in-plane magnetic field angle in 2D $d$-wave
superconductors}

\author{G. R. Boyd}
\author{P. J. Hirschfeld}
\affiliation{Department of Physics, University of Florida,
    Gainesville, FL 32611, USA}
\author{I. Vekhter} \author{A. B. Vorontsov}\altaffiliation{
Present address: Dept. of Physics, University of Wisconsin,
Madison, WI} \affiliation{Department of Physics and Astronomy,
Louisiana State University, Baton Rouge, Louisiana 70803 USA   }

\begin{abstract}
Experiments on several novel superconducting compounds have
observed oscillations of the specific heat when an applied
magnetic field is rotated with respect to the crystal axes. The
results are commonly interpreted as arising from the nodes of an
unconventional order parameter, but the identifications of nodal
directions are sometimes controversial. Here we show with a
semiclassical model calculation that when the magnetic field
points in the direction of the nodes, either minima or maxima can
occur in the specific heat depending on the the temperature $T$
and the magnetic field $H$. An
inversion of the angular oscillations takes place with respect to
those predicted earlier at low temperature by the nodal
approximation.
 This result, together with the argument that the inversion
takes place based on an approximation valid at moderate fields,
indicates that the inversion of  specific heat oscillations is
an intrinsic feature of nodal superconductors.
\end{abstract}
\pacs{}

\maketitle

\section{Introduction}

Initial information on the symmetry of the order parameter in
newly discovered superconductors is often
provided by power laws in the temperature dependence of
thermodynamic and transport properties \cite{NEHussey:2002}.
However, the exact nodal structure cannot  be determined by these
techniques because the same exponents may correspond to  order
parameters with different symmetries. Phase-sensitive experiments
like the tricrystal experiments performed on cuprates
 {\cite{CCTsuei:2000}} are the most definitive, but are
technically challenging and require high quality samples.  A
simpler technique, proposed in Ref. \onlinecite{IVekhter:1999R},
which is not phase sensitive, but provides information on the
distribution of nodes  on the Fermi surface, is a measurement of
the specific heat in the presence of a magnetic field,  {$\bf H$},
which is rotated with respect to the crystal axes  of the sample.
It was shown in that work that, if the effect of the
superflow
 {due to vortices} on the quasiparticle spectrum was treated
 semiclassically, via the Doppler energy shift \cite{GEVolovik:1993,CKubert:1998SSC}, the low-temperature
specific heat of a superconductor with line nodes acquires an
oscillatory dependence on the field orientation. In particular,
the low-temperature  specific heat coefficient $\gamma\sim
T\sqrt{H}$ was shown to have minima when the magnetic field points
in the direction of the nodes in the order parameter, and maxima
for the field along the antinodes, where the gap is maximal.
Ref.~\onlinecite{IVekhter:1999R} and its extension,
Ref.~\onlinecite{IVekhter:2001}  made use of  {two additional
approximations. First, they replaced the Doppler energy shift of a
quasiparticle with momentum $\k$ near one of the nodes by its
value at the node, $\k=\k_n$.  This so-called nodal approximation
had been shown to work well compared to the full semiclassical
evaluation in Ref. \onlinecite{CKubert:1998SSC}, providing $T, E_H
\ll \Delta_0$. Here $\Delta_0$ is the maximum gap over the Fermi
surface, and $E_H\sim \Delta_0\sqrt{H/H_{c2}}$, where $H_{c2}$ is
the upper critical field, is a magnetic energy scale (see Sec.
\ref{semiclass}). Second, making predictions for low temperatures
and fields, $T, E_H \ll \Delta_0$, the authors of
Refs.~\cite{IVekhter:1999R,IVekhter:2001} used a form of the
order parameter linearized in the vicinity of the nodal points.}

By now there have been several experimental tests of these
ideas~\cite{TPark:2003,TPark:2004,HAoki:2004,KDeguchi:2004,KDeguchi:2004JPSJ,TSakakibara:2007,KYano:2008},
and the observations have been generally consistent with
theoretical expectations. At the same time, the assignment of the
nodal directions in several materials remains controversial. In
CeCoIn$_5$ the measurements of the anisotropy of the specific
heat~\cite{HAoki:2004} and the thermal
conductivity~\cite{KIzawa:2001,YMatsuda:2006} appear to give
contradictory results for the gap structure. In Sr$_2$RuO$_4$ the
specific heat oscillations were observed to {\it invert} as the
temperature and field was varied, i.e. the minima and maxima as a
function of angle changed
places~\cite{KDeguchi:2004,KDeguchi:2004JPSJ}.

Such an inversion was never found in earlier theoretical
calculations within the semiclassical approach, even though
it was found in theoretical work employing other
techniques, see below. This clearly poses a problem: if the
technique is to be useful as a way to ``map" out the nodal
structure, it is necessary to be able to predict if and when such
inversions will occur; otherwise the nodes in the order parameter
may be assigned to incorrect locations in momentum space.

Numerical solution of the Bogoliubov-deGennes equations yielded an inversion in the
anisotropy of the density of states (DOS), $N(\omega,\theta)$,
between the field applied in the nodal and antinodal
directions~\cite{MUdagawa:2004}, but this was argued to reduce,
rather than invert, the specific heat anisotropy. Recently
Vorontsov and Vekhter~\cite{ABVorontsov:2006,ABVorontsov:2007}
considered the limit $H_{c1}\ll H\leq H_{c2}$ by extending the
method of Brandt, Pesch and Tewordt\cite{BPT:1967,WPesch:1975},
and found an agreement with semiclassical method at low
$T,H$, but an inversion of the {specific heat} oscillations over
a large part of the {$T-H$} phase diagram. Since the
approximation they used was tailored for moderate to high fields,
their determination of location (or, indeed, the existence) of the
inversion line could be questioned.

In this paper we demonstrate that this inversion is a generic
feature of the specific heat in unconventional superconductors. We
revisit the semiclassical approach to the in-plane specific heat
oscillations of a quasi-two dimensional $d$-wave superconductor,
but relax the approximations which led in Refs.
\cite{IVekhter:1999R,IVekhter:2001,volovik2} to simple
analytical forms for $C(T;\theta)$ at low temperatures.  We are
thus forced to do a numerically demanding evaluation of the
entropy and specific heat, which involves a 2D $\k$-summation and
a 2D averaging over a vortex lattice unit cell. Our model is not
 restricted to low temperatures, and is valid (within limits
discussed below) in the low field regime $H_{c1} \lesssim H\ll
H_{c2}$.  It may thus  be considered to  complement the results of
Refs. \cite{ABVorontsov:2006,ABVorontsov:2007}. We find that
the anisotropy of the specific heat is inverted as the temperature
is increased. This result demonstrates that the inversion
phenomenon is robust across the phase diagram for unconventional
superconductors, and provides an important caveat to the
interpretation of the stationary points in specific heat
oscillations.

\section{The Semiclassical Approach}
\label{semiclass}

In superconductors with line nodes, the vortex core contribution
to the low-energy density of states is smaller than that from the
quasiparticles outside the core region \cite{GEVolovik:1993}. In
systems with short coherence length, such as cuprates and heavy
fermion materials, the dominance of the extended quasiparticle
states is even more pronounced since  {the spacing of the energy
levels in the vortex core is large, and only few such levels (if
any exist at all) are occupied at low temperature}. Bulk
quasiparticles in the vortex state are excited from the pair
condensate moving with the superflow around each vortex; hence the
effect of applied magnetic field can be simply described by
Doppler-shifting the spectrum of extended quasiparticle states
according to the local value of the superfluid velocity, $\vs
(\rr)$. This approximation is valid at $H\ll H_{c2}$, when the
vortices are far apart and $\vs (\rr)$ varies slowly on the scale
of the superconducting coherence length. Using this approach,
Volovik predicted~\cite{GEVolovik:1993} that the density of states
at the Fermi level in the vortex state of a superconductor with
line nodes varies as $N(\omega=0;H)\propto\sqrt H$. Specific heat
measurements on high-T$_c$ cuprates verified the $\sqrt{H}$ field
dependence\cite{KAMoler:1994,KAMoler:1997,YWang:2001}, which
played an important role in the identification of $d$-wave
symmetry.  It should be noted, however, that the result holds for
any gap with line nodes, including, for example, some $p$-wave
states. The $\sqrt{H}$ dependence is modified by the presence of
disorder to $H \log H$ as shown by K\"{u}bert and
Hirschfeld\cite{CKuebert:1998}, but remains nonlinear and
qualitatively similar to the pure case.

The semiclassical approximation provides a conceptually
transparent and tractable approach to the effect of the vortex
lattice on low-lying quasiparticle states. Its validity has been
questioned by several authors, who attempted a more accurate
description of the effect of both applied field and supercurrents
on the quasiparticle spectrum. Franz and
Te{\v{s}}anovi{\'c}~\cite{MFranz:2000} introduced a singular gauge
transformation that takes into account both the supercurrent
distribution and the magnetic field, and mapped the full
quantum-mechanical problem onto that of nodal Dirac fermions
interacting with effective scalar and vector potentials that are
periodic in the unit cell of the vortex lattice. They discovered
significant differences from the quasiclassical theory in the
quasiparticle excitations at very low energy, as is to be expected
since the semiclassical approach works only for large quantum
numbers.  We do not discuss the details of this breakdown here,
since it was done by  Knapp et al.
\cite{DKnapp:2001,DKnapp:2001E}, who explicitly compared the
quantum-mechanical and semiclassical results for the density of
states, and found that small differences between the two
approaches begin to appear below a crossover scale which is
exponentially small in the Dirac cone anisotropy $v_F/v_2$. Here
 $v_F$ is the Fermi velocity, and $v_2\equiv d\Delta/dk_\|$ is the
gap ``velocity'' at the node ($k_\|$ is the momentum component
along the Fermi surface). Since in real materials this ratio is
large, and in real samples the presence of impurity scattering and
the disorder in the vortex lattice smear out the energy structure
on small scales, we assume that for  purposes of comparison with
the measurements, the semiclassical description is adequate.

Dahm et al. \cite{TDahm:2002} investigated in detail the
comparison of the  simple single vortex Doppler shift approach
with the solution of the quasiclassical microscopic Eilenberger
equations using several techniques 
including the Brandt, Pesch, and Tewordt
approximation~\cite{BPT:1967,WPesch:1975}. At distances from the
vortex of the order of the coherence length, the Doppler shift
method is quantitatively inadequate because of core state
contributions or scattering resonances, but these effects have
little qualitative impact on the calculation of thermodynamic
properties. We therefore proceed with the simplest Doppler-shift
analysis for a single vortex unit cell, in order to make the
qualitative point which is the main result of this work.

For concreteness, we consider  { a quasi-two dimensional
superconductor} with $d$-wave symmetry. This situation is most
closely realized in cuprates, possibly the heavy fermion 115
compounds\cite{RMovshovich:2001,KIzawa:2001},  and potentially
also the newly-discovered oxypnictide
materials\cite{Kamihara:2008}. We  {consider the field applied
parallel to the conducting plane, ${\bf H}\parallel ab$, and
assume that the system is sufficiently three-dimensional, so that
at the fields of interest the structure of the mixed state
resembles the Abrikosov vortex lattice penetrating a stack of
weakly coupled 2D planes \cite{JClem:1990}. In this case the
vortex superflow field ${\bf v}_s$ is three dimensional, different
on different planes within the vortex unit cell. For weak
interlayer coupling we consider a circular in-plane Fermi surface,
and therefore account only for quasiparticles with momenta ${\bf
k}\parallel ab$.} Deviations from Fermi surface isotropy and
effects of  {multiple Fermi surface sheets} have been studied e.g.
in Refs. \cite{Graser:2008} and \cite{YTanaka:2003}, but do not
affect the qualitative conclusions we wish to draw here.

The single-particle Green's function in the presence of a
superflow velocity field $\vs$ is obtained by Doppler shifting the
quasiparticle states with energy $\omega$ and momentum
$\k$\cite{CKubert:1998SSC}  {(we use units with $\hbar=k_B=1$)},
\begin{equation}
G(\k,\rr, \omega_n)=-\frac{(i\omega_n-\vs
(\rr)\cdot\k)\tau_0+\Delta_{\widehat\k}\tau_1+\xi_\k \tau_3}{
(i\omega_n-\vs (\rr)\cdot\k)^2-\xi_\k^2-\Delta_{\widehat\k}^2}\,,
    \label{eq:GF}
\end{equation}
where $\omega_n$ is the fermionic Matsubara frequency, $\xi_{\bf
k}$ is the band energy measured with respect to the Fermi level,
$\tau_i$ are  {Pauli} matrices in particle-hole space,
 {and $\Delta_{\widehat\k}=\Delta_0\cos 2\phi$, with $\phi$
the azimuthal angle on the Fermi surface.}

We consider the magnetic field,  {$\bm H$}, applied in the ab
plane,  {and approximate the superflow by that of an anisotropic
3D superconductor in the London model. The contours of constant
$\vs$ are then elliptical due to anisotropy of the penetration
depth, $\lambda_{ab}\neq\lambda_c$. After rescaling the $c$-axis
by {\textbf{$z'=z {\lambda_{ab}}/{\lambda_{c}}$}}, the superflow
is cylindrically symmetric and the Doppler shift for
quasiparticles at the Fermi surface is given
by~\cite{IVekhter:1999R,IVekhter:2001} }
\begin{equation}
 \vs (\rr)\cdot\k_F = \frac{\hbar k_F}{2 m r}
 \sin(\psi)\sin(\theta-\phi)\,,
 \label{eq:vs}
\end{equation}
where $\psi$ is the winding angle of the superfluid velocity in
real space, $\theta$  is the angle  { between $\bm H$ and the
a-axis, and $r$ is the rescaled distance from the center of the
vortex}. Note that we explicitly exclude from consideration
quasi-1D Fermi surfaces such as those shown by Tanaka et
al.\cite{YTanaka:2003} to lead to $C(H;\theta)$ oscillation
inversion at low temperatures.

From Eqs. (\ref{eq:GF})  the local density of states is
\begin{eqnarray}
  N (\omega,\rr)&=& -\frac{1}{2\pi}{\rm Im} \sum_{\k}
 {\rm Tr }\,G( \k,\omega-\vs (\rr)\cdot \k)\,
  \label{eq:ldos}\\
    &\simeq& N_0 {\rm Re }\int_0^{2\pi}\frac{d\phi}{2\pi}
    \frac{|\omega-\vs (\rr)\cdot \k_F|}{\sqrt{(\omega-\vs (\rr)\cdot
    \k_F)^2-|\Delta_0(\phi)|^2}}\,.
   \nonumber
\end{eqnarray}
Here $N_0$ is the normal state density of states. The  {net DOS
per volume} is found by spatially averaging $N(\omega,\rr)$ over a
unit cell of the vortex lattice containing one flux quantum,
$\Phi_0$.  {After rescaling the $c$-axis the unit cell area
increases by a factor $\lambda_{c}/\lambda_{ab}$, and therefore
flux quantization dictates that the quasiparticles now experience
the effective field $H^\star=(\lambda_{ab}/\lambda_c)H$.} In the
new coordinates
 {the radius and the area of such a cell are
$R_H=\sqrt{\Phi_0/\pi H^\star}$ and $A_H=\pi R_{H}^2$
respectively.} Introducing polar coordinates,
$\rr=R_H(\rho\cos\psi,\rho\sin\psi)$, we find for the average
field-dependent density of states
\begin{equation}
  N(\omega,\mathbf{H},T)={1\over \pi}
\int_0^1\rho d\rho\int_0^{2\pi}d\psi\, N(\omega,\rr)\,.
\label{eq:dos}
\end{equation}

 {The density of states depends on the angle between the field
and the crystal axes. At low energies the dominant contribution to
the local DOS is from the near-nodal regions,
$\phi\sim\phi_n=\pi/4\pm n\pi/2$, with $n=0\ldots 3$, and
oscillations in $N(\omega,\mathbf{H},T)$ appear since the Doppler
shift at a given node vanishes when the field is aligned with that
node. In the nodal approximation with linearized order parameter,
the analytical form of the density of states was
obtained~\cite{CKubert:1998SSC,IVekhter:1999R,IVekhter:2001}, see
next Section; however, as pointed out before, this approximation
does not give the reversal of the anisotropy (see below).}


In all of the following we carry out the full summation over the
momenta and the real space coordinates. We use the density of
states to compute the entropy,
\begin{eqnarray}
S&=&-2 \int_{-\infty}^{\infty} d\omega N(\omega)
[(1-f(\omega))\log(1-f(\omega))\nonumber\\&&  + f(\omega)\log
f(\omega)]\,, \label{entropy}
\end{eqnarray}
where $f(\omega)=(\exp(\omega/T)+1)^{-1}$ is the Fermi function.
To obtain the  the heat capacity at constant volume we
differentiate,
\begin{equation}
C_V (T,{\bf H})= T\left(\frac{\partial S}{\partial
T}\right)_{H,V}\,\label{spht}\,.
\end{equation}
At low $T$ and $H$, when the gap varies weakly with temperature,
the temperature derivative acts only on the Fermi function in
Eq.(\ref{entropy}), and the specific heat is given by the simple
 form
\begin{equation}
C_V(T,{\bf H})\approx  \frac{1}{2} \int_{-\infty}^{\infty}
d\omega\,N(\omega,{\bf{H}}) \frac{\omega^2}{T^2} {\rm
sech}^{2}\frac{\omega}{2T} \,. \label{eq:lowspht}
\end{equation}
 {Here we consider the full range of temperatures below the
transition, and hence evaluate the specific heat from
Eq.~(\ref{spht}). We approximate the temperature dependence of the
order parameter, $\Delta_\k(T)$, }according to the BCS
weak-coupling ansatz appropriate to a circular Fermi
surface\cite{Einzel}
\begin{equation}
\Delta(T)=\Delta_0 \cos(2 \phi)\, \tanh \left(\frac{\pi
T_c}{\Delta_0}
\sqrt{\frac{4}{3}\frac{8}{7\zeta(3)}(\frac{T_c}{T}-1) } \right),
\end{equation}
where $\Delta_0=2.14T_c$ is the gap maximum.

 {In the absence of impurity scattering,} there are two
important low energy energy scales in the problem: the temperature
$T$ and the magnetic energy, or typical Doppler shift $E_H\equiv
v_F/R_H$.  To satisfy the requirements of the semiclassical
approach, we  {consider only} $E_H\ll \Delta_0$, but temperature
to vary over the entire range $T<T_c$.

\section{Density of states in planar field}
\label{sec:DOS}

In the nodal approximation of Ref. \onlinecite{IVekhter:1999R},
the quasiparticle momentum $\k$ is replaced by its value  $\k_n$
at each of the 4 nodes, which are then summed over.   The
 {residual density of states at the Fermi level},
Eq.~(\ref{eq:dos}) then becomes
\begin{equation}
\frac{N(0,{\bf H})}{N_0} = {2\sqrt{2}\over \pi}{E_H\over \Delta_0}
\beta (\theta)\, ,\label{resdos}
\end{equation}
where the angular variation is given by $\beta(\theta)={\rm
max}\,(|\sin\theta|,|\cos\theta|)$.  Eq.~(\ref{eq:lowspht}) then
yields the linear in temperature specific heat at low $T$ with the
slope
\begin{equation}
     {\lim_{T\rightarrow 0}\frac{C_V(T;{\bf H})}{T}\simeq \frac{4\sqrt{2} \pi}{3} {E_H\over
    \Delta_0} \beta(\theta)}\,.
    \label{CVlinearterm}
\end{equation}
 {At low energies, $E_H,\omega\ll\Delta_0$, the density of
states in the nodal approximation with linearlized order parameter
takes the form~\cite{IVekhter:1999R,IVekhter:2001}
\begin{equation}
  \frac{N(\omega, {\bf  H},
  T)}{N_0}=\frac{1}{2}\left[
  \frac{E_1}{\Delta_0}F\left(\frac{\omega}{E_1}\right)
  +
  \frac{E_2}{\Delta_0}F\left(\frac{\omega}{E_2}\right)
  \right]
  \,,
  \label{eq:dosnodalfreq2}
\end{equation}
where $E_1=E_H|\sin(\pi/4-\theta)|$,
$E_2=E_H|\cos(\pi/4-\theta)|$, and the scaling function $F$ is
given by \cite{CKubert:1998SSC}
\begin{equation}
  F(y)=
  \begin{cases}
    y\left[1+1/(2y^2)\right], & \text{if $y\geq 1$;}\\
    \left[(1+2y^2)\arcsin y + 3y\sqrt{1-y^2}\right]/y\pi, &
    \text{if $y\leq 1$.}
  \end{cases}
  \label{eq:Fdos}
\end{equation}
} Note that in the limit $y\rightarrow \infty$, $F(y)\rightarrow
y$, such that $N(\omega,{\bf H},T)$ in (\ref{eq:dos}) recovers the
{\it isotropic} low-$\omega$ $d$-wave density of states
$\omega/\Delta_0$. Thus in the nodal approximation with
linearized order parameter the specific heat oscillations are
washed out at higher temperatures, but the method cannot generate
specific heat oscillation inversions.

\begin{figure}[h]
\includegraphics[width=90mm]{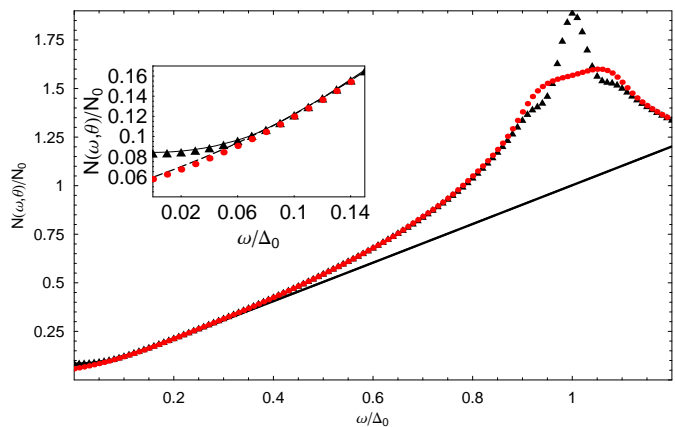}
\caption{(Color online)  Comparison of the nodal approximation
Eq.(\ref{eq:dosnodalfreq2}) for the density of states $N(\omega,{\bf
H})$ versus the full density of states at $T=0.001T_c$ and
$E_H=0.2 T_c$.  solid line: nodal approximation for $\bf H$
along antinode. Dashed line: nodal approximation for $\bf H$ along
node. Red circles: full evaluation for $\bf H$ along node. Black triangles: full evaluation for $\bf H$ along antinode. The
insert magnifies the low-frequency range. }\label{nodalcompare}
\end{figure}

In Fig.~\ref{nodalcompare}, we now compare the density of states
of a $d$-wave superconductor obtained from a complete evaluation
of Eq.~(\ref{eq:dos}) with the nodal approximation with linearized
order parameter Eq. \ref{eq:dosnodalfreq2}.  In agreement
with all previous work the density of states for field in nodal
and antinodal directions is strongly anisotropic for
$\omega\lesssim E_H$, and the anisotropy is washed out at higher
energies. However, the difference between $N(\omega,{\bf H},T)$
for the two directions of the field reappears at energies of order
$\Delta_0$, the energy scale absent in the versions of the
calculation with linearized gap.

For the field along the antinode $\bm H\|\k_{an}$
($\theta=0,\pi/2$...) the density of states continues to be
sharply peaked at $\Delta_0$, as in the absence of the field. This
peak is largely due to the quasiparticles moving along the field,
which do not experience the Doppler shift and see the full maximal
gap. In contrast, for the field along the node, $\bm H\|\k_n$
($\theta=\pi/4, 3\pi/4,\dots$), the DOS has broad features
around $\Delta_0\pm E_H$. As a result, the density of states for
the field along the nodal direction begins to exceed that for
field along the antinode at the lower shoulder, and the DOS
anisotropy is inverted, relative to that at low energies.

We emphasize that the absence of the structure near the gap
edge in $N(\omega,{\bf H},T)$ in previous work is a consequence of
gap linearization, and not the nodal approximation. We observe
features similar to those depicted in Fig.~~\ref{nodalcompare}
when replacing $\k\rightarrow \k_n$ in the Doppler shift, but
keeping the full variation of the order parameter around the Fermi
surface. Of course, the nodal approximation is not accurate at the
energies of order $\Delta_0$, and hence we continue with the full
evaluation of the DOS.

The anisotropy in the density of states that we computed
differs from that found in
Refs. \cite{ABVorontsov:2006,ABVorontsov:2007}, where
noticeable inversion of the anisotropy in $N(\omega,{\bf
H},T)$ occurred already at relatively low energy, as a result of
the competition between Doppler shift and vortex scattering.
Remarkably,  we find that even within the semiclassical method,
when the lifetime remains infinite and  the scattering on the
vortices in neglected, the anisotropy in the density of state is
still reversed, albeit at higher energies of order of the gap
maximum.

The possibility of an inversion of the specific heat anisotropy is
clear from Fig. \ref{nodalcompare}. In
Eq.~(\ref{eq:lowspht}) the density of states is convoluted with
the temperature-dependent weighting function, peaked around
$\omega\simeq 2.5 T$. At a given field, as the temperature is
increased, more weight in the kernel shifts to higher energies,
where the anisotropy in the DOS is opposite to that at low
$\omega$. Whether an inversion then occurs at higher $T$ depends
primarily on whether the kernel has sufficient weight in the
exponential tail at energies $\omega\simeq \Delta_0-E_H$.  Since
the DOS anisotropy is small in magnitude above this crossing
energy, the specific heat must be calculated numerically.

\begin{figure}[h]
\includegraphics[width=90mm]{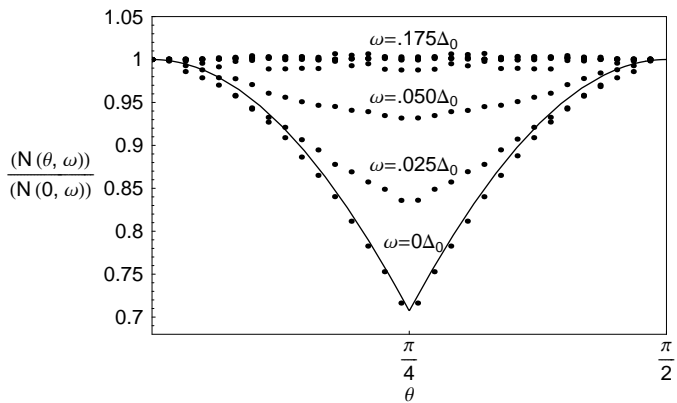}
\caption{DOS $N(\theta,\omega)$ vs. magnetic field angle $\theta$
at $E_h=0.2 T_c$ and $T=0.05T_c$ for equally spaced frequencies,
0.025$\Delta_0$ apart, from $\omega=$ 0 to 0.175$\Delta_0$.  The
solid line is the result of the nodal approximation for the
$\omega=0$ DOS from Eq. (\ref{resdos}). }\label{dos_oscil}
\end{figure}

In Figure 2 we plot the density of states as a field sweeps
through the ab-plane. We clearly see that for low frequency the
node and minimum in the oscillations coincide, and that at higher
frequencies this is no longer the case. The eventual inversion of
this pattern results in the angular density of states having a
maximum at the gap node for higher frequencies.

\section{Specific heat oscillations}

\begin{figure}[h]
\includegraphics[width=60mm,angle=-90]{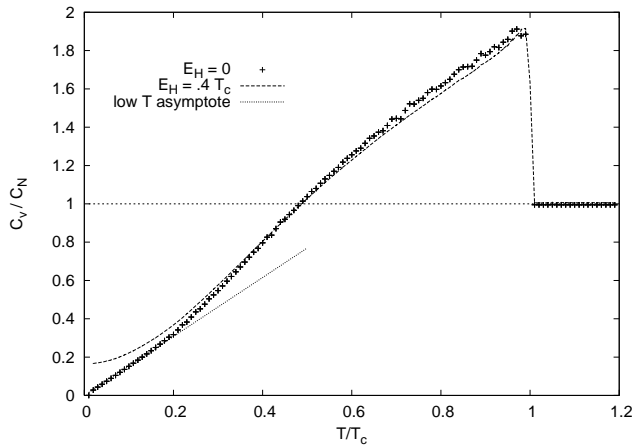}
\caption{Temperature dependence of the specific heat (at constant volume) normalized to the specific heat in the normal state
$C_V/C_N$, where $C_N=\gamma_N T$. We show both the exact (symbols), and the asymptotic
low-$T$ (dotted line) behavior in the absence of the field,
and compare it to the numerically determined $C_V/C_N$ at  $E_H=0.4
T_c$  (dashed line). There is no reduction of $T_c$ with
field in our approach.} \label{CvsT}
\end{figure}

Numerical differentiation of the entropy is computationally
intensive due to the high accuracy required in finding $S$
in Eq.~(\ref{entropy}). To illustrate the precision of our
calculations in Fig.~\ref{CvsT}, we show the numerically
evaluated specific heat at $H=0$ and at $E_H=0.4T_c$ for
the field along the a-axis {over a wide temperature range both
below and above $T_c$.} The normal state specific heat above $T_c
$ is ${C_N}=\gamma_N T \equiv{2N_0 \pi^2 T}/{3}$, where $N_0$ is the Fermi level
DOS per spin.

To test our numerical evaluations we  {first check the numerical
results against} the asymptotic low-temperature specific heat in
zero field. For our model of a 2D $d$-wave superconductor with a
circular Fermi surface  {at $T\ll T_c$} we find
\begin{eqnarray}
\hspace*{-100pt}&&\frac{C_V}{ N_0 T} \approx \gamma_0
\frac{T}{T_c}\,,
\\
&&\gamma_0 =\frac{T_c}{\Delta_0} \int_0^{\infty} \frac{x^3 \
dx}{\cosh^2(x/2)} = \frac{ T_c}{\Delta_0} 18 \zeta(3) \approx
10.11\,,\ \qquad
\end{eqnarray}
 {which agrees with the numerically determined slope.} The
Volovik effect is manifested  {in Fig.~\ref{CvsT}} in the finite
offset of $C/T$ (Eq.~(\ref{CVlinearterm}) in the presence of
magnetic field $E_H=0.4T_c$ .

\begin{figure}[h]
\includegraphics[width=80mm]{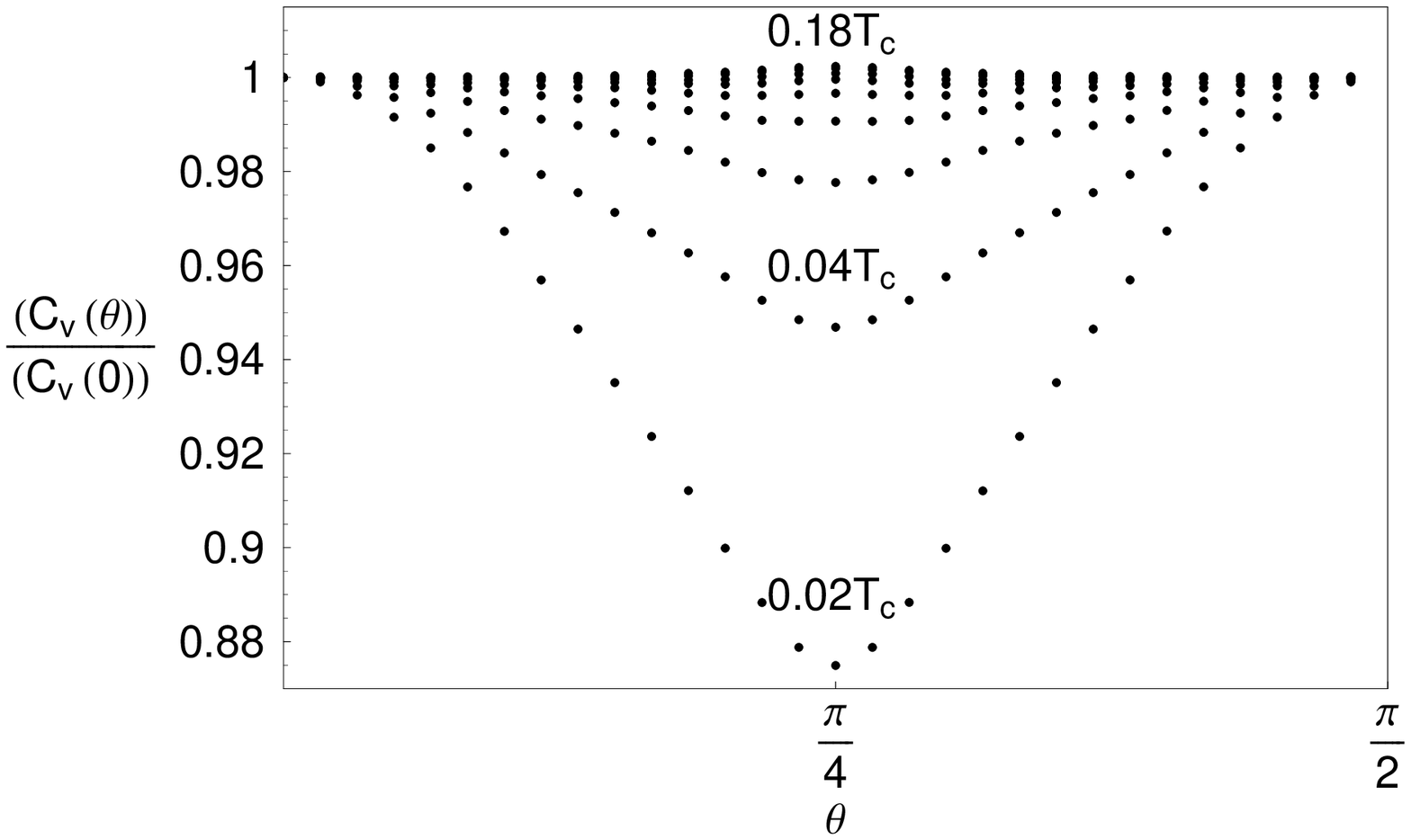}
\includegraphics[width=80mm]{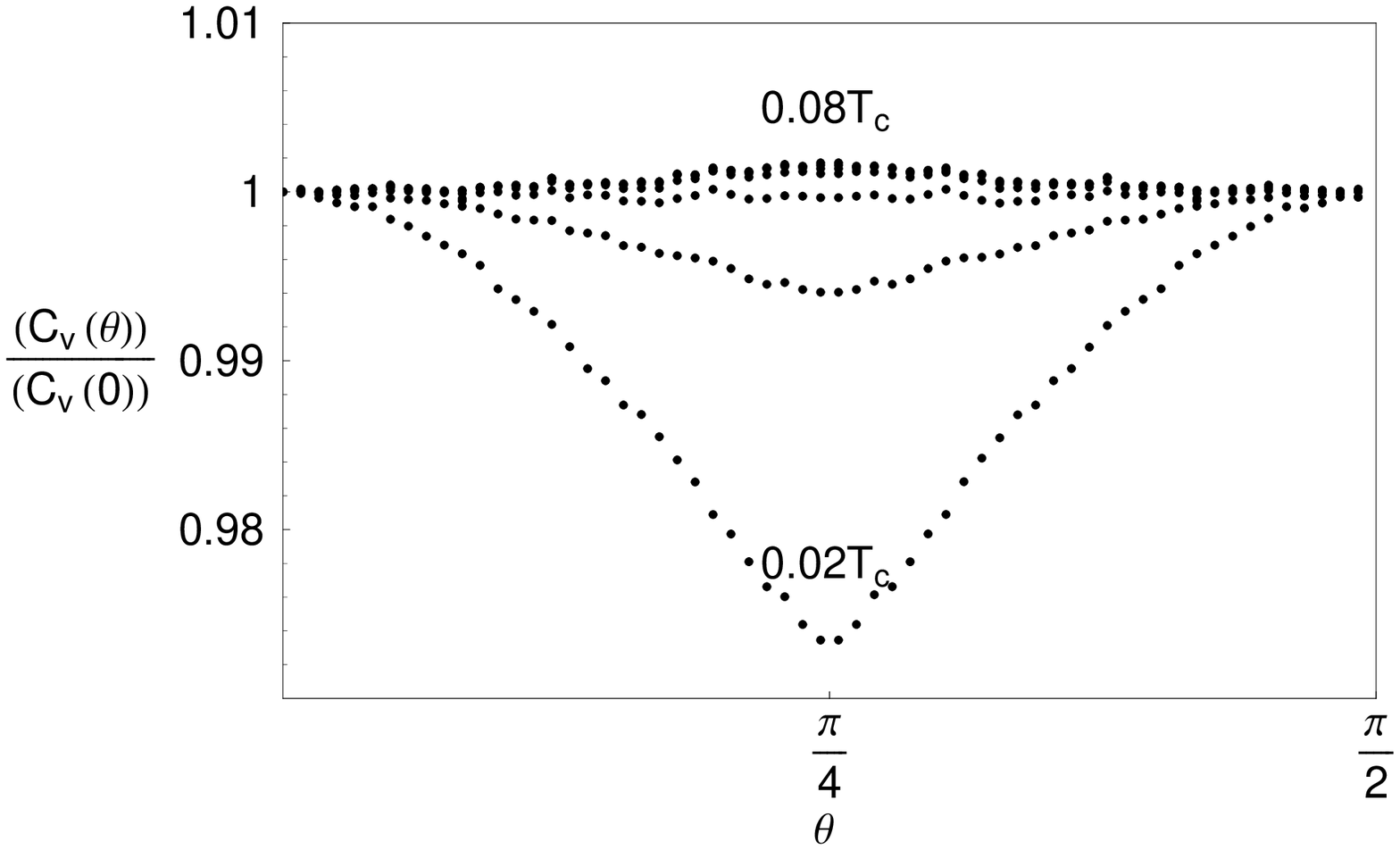}
\caption{ Upper panel: ${C(T,{\bf H})}/{C(T,{\bf \theta=0})}$ vs.
$\theta$ for a set of equally spaced temperatures, every 0.02$T_c$,
from 0 to 0.18$T_c$, $E_h=0.2T_c$.  Lower panel: same for a set of
equally spaced temperatures from 0 to 0.08 $T_c$ every 0.02$T_c$,
$E_h=0.05T_c$. } \label{C_vs_theta}
\end{figure}

We proceed to evaluate the specific heat for fixed $E_H$ and
several temperatures as a function of the field angle $\theta$.
Fig.~\ref{C_vs_theta} shows that the  {inversion of the DOS
anisotropy found in Sec.~\ref{sec:DOS} indeed are sufficient to
lead to the inversion of the specific heat oscillations a
characteristic temperature, $T_{inv}$. In Fig.~\ref{C_vs_theta}
the fourfold oscillations are clearly visible at low $T$, and
minima occur for $\bf H$ along nodal directions as anticipated.
However, at higher temperatures, an inversion in the pattern of
oscillations is evident. In Fig. \ref{crossover} we examine the
anisotropy by plotting the difference between the specific heat
for the field along the nodal and the antinodal directions, which
identifies the temperature at which inversion occurs.

 Since in our approach the inversion is due to the
sensitivity of the specific heat to the changes in the DOS
within energy range $\sim E_H$ of the gap edge, increasing
both the magnetic field and the temperature  initially enhances
the amplitude of the inverted oscillations. Increasing the field
brings the anisotropy inversion down in energy  in
Fig.~\ref{nodalcompare}; raising the temperature increases the
contribution  of the high energy regions in C/T. While the
amplitude of the inverted (relative to those at $T=0$)
oscillations in $C_V$
is small, it is of the same order of magnitude as that observed
experimentally
\cite{TPark:2003,HAoki:2004,KDeguchi:2004,TSakakibara:2007,KYano:2008},
and found theoretically for a quasi-two-dimensional system at
moderate fields~\cite{ABVorontsov:2006,ABVorontsov:2007}. Of
course, when $H$ ($T$) approaches the upper critical field (the
transition temperature) respectively, the gap in the spectrum
closes, and the oscillations vanish; we cannot, however, reliably
comment on the evolution of the anisotropy in this regime within
the semiclassical method. At the same time it follows from our
analysis that the amplitude of the inverted oscillations has a
maximum at intermediate fields and temperatures, also in agreement
with Refs.~\cite{ABVorontsov:2006,ABVorontsov:2007}.
Therefore our results connect well with those obtained by a
different technique.

\begin{figure}[t]
\includegraphics[width=80mm]{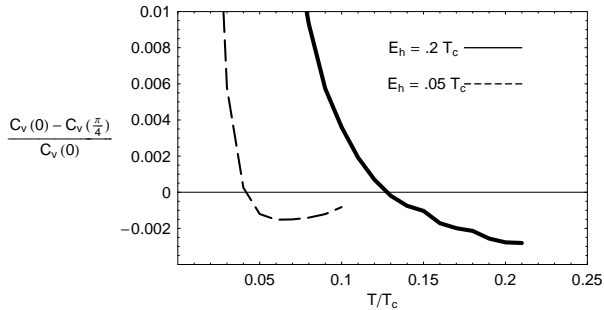}
\caption{Amplitude of specific heat oscillation defined by
difference of $C_V$ between node and antinode,
$[{C_V(0)-C_V(\frac{\pi}{4})}]/{C_V(0)}$
 for $E_H=0.2 T_c$ (solid line) and $E_H=0.05T_c$ (dashed line).  }\label{amplitude}
\label{crossover}
\end{figure}

\section{Conclusions}

In this paper we have calculated the specific heat of a
 {two-dimensional} $d-$wave superconductor in an external
magnetic field using the semiclassical treatment of the effect of
the vortex lattice on the quasiparticle spectrum.   {In contrast
to previous work utilizing the nodal approximation with linearized
order parameter, we carried out a full numerical evaluation of the
density of states and the entropy for a wide range of fields and
temperatures. Our main finding is that the {\em sign} of the
oscillations of the specific heat as a function of the field
orientation, i.e. the difference between $C_V$ for the field along
a nodal direction and along the antinode, depends on the
temperature and field strength. We confirmed that at low
temperatures and fields the specific heat has a minimum when the
field is along a nodal direction. However,  as $H$ and $T$ are
increased, minima of the specific heat begin to occur for the
field along the gap maxima, i.e. an inversion of the oscillation
pattern occurs. Note that while we considered a  system with
well-defined quasiparticle states,  it is reasonable to believe
that scattering due to impurities or vortex lattice disorder will
merely smear the anisotropy on both sides of the inversion line.


Our calculations   provide a bridge connecting the semiclassical
theory at low fields $H<<H_{c2}$ with the results of the extended
Brandt-Pesch-Tewordt
approximation~\cite{ABVorontsov:2006,ABVorontsov:2007}, where the
inversion was first found. The latter approach is in principle
valid at $H\lesssim H_{c2}$, but has been shown to provide
remarkably good agreement with semiclassical predictions down to
low fields, up to log corrections~\cite{ABVorontsov:2007}.  To
compare the results of the two approaches explicitly, we
extend our calculation to higher fields, account for the field
dependence of the gap amplitude via
$\Delta(T,H)=\Delta(T)\sqrt{1-H/H_{c2}}$, and determine the
inversion line. This provides a direct comparison with the results
of Ref.~\onlinecite{ABVorontsov:2007}, where the same field
dependence was assumed for a circular Fermi surface identical to
that considered above. In Fig. \ref{PhaseD} we plot the
approximate crossover scales for both the Brandt-Pesch-Tewordt and
semiclassical theories. There is a remarkably good
qualitative agreement for the behavior of the inversion line up to
moderate fields, where the ranges of validity of the two
approaches may reasonably be assumed to overlap. This establishes
a phase diagram for when the specific heat is expected to have
minima or maxima at the gap nodes.
\begin{figure}[h]
\includegraphics[width=85mm]{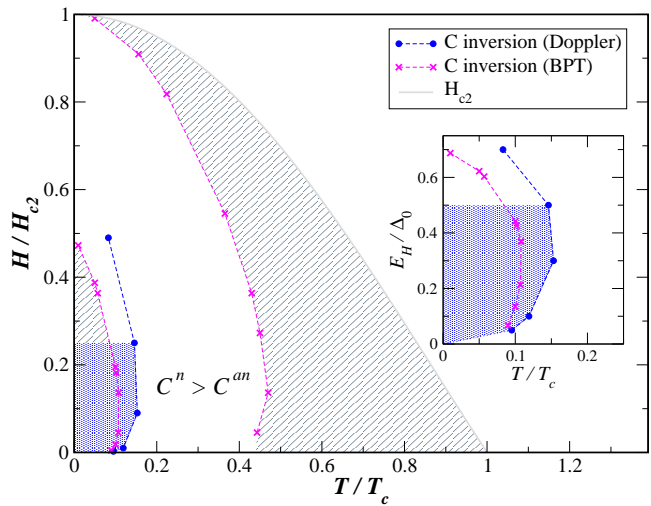}
\caption{ Phase diagram  $H/H_{c2}$ vs. $T/T_c$ for specific heat
oscillations of a $d$-wave superconductor in a magnetic field. The
 shaded regions indicate that the specific heat $C(T,{\bf H})$ has a minimum
 when the field ${\bf H}$ points in the nodal direction, whereas the white
 regions indicate inverted oscillations, where minima correspond
 to field along antinode.  The gray shaded regions are
results from Ref. \onlinecite{ABVorontsov:2007} using the
Brandt-Pesch-Tewordt framework, whereas the blue regions represent
the region of the $H-T$ plane where minima
correspond to nodes within the semiclassical theory. The phase
diagram comparing both approaches was determined using the
assumption $H/H_{c2}= (E_H/\Delta_0)^2$, valid up to a prefactor
of order unity. Insert: blowup of the low $T$ and $H$ region.
Note the field scale is given in terms of $E_H/\Delta_0$ for
easier comparison with other results in this work. }
\label{PhaseD}
\end{figure}
Taken together, these results strongly suggest that the inversion
of the specific heat oscillations with temperature is a general
feature for all nodal superconductors, and establish the
approximate location of the inversion line.

Consequently, identification of the nodes in the
gap via the specific heat measurements is not as straightforward
as the original semiclassical results suggested, and depends on
where the experiment is done in the field-temperature plane. While
the inversion of the oscillations in related experiments on the
anisotropy of the heat conductivity of nodal superconductors has
been well
established~\cite{HAubin:1997,TWatanabe:2004,YMatsuda:2006,YKasahara:2008},
an inversion in the heat capacity measurements has not yet been
searched for systematically. Experimental identification of the
inversion line in the $T$-$H$ plane would be an important step
towards further establishing the heat capacity measurements as the
method of choice for determining the nodal directions in the bulk,
and our paper provides theoretical foundation for such a search.

\section*{Acknowledgments}

This work was supported in part by DOE DE-FG02-05ER46236 (P.J.H.
and G.R.B.), the Louisiana Board of Regents (I.V. and A.B.V.), and
DOE DE-FG02-08ER46492 (I.~V.). This work was started at the
Institut d'\'Etudes Scientifiques de Carg\`ese, and supported
there in part by the I2CAM via NSF grant DMR 0645461. GRB and PJH
acknowledge useful discussions with S. Graser.


\end{document}